\documentclass[10pt,letterpaper,twocolumn]{article} 

\usepackage{ol2}
\usepackage[draft]{hyperref}
\usepackage{amsmath}
\usepackage{amssymb} 
\usepackage{bm}
\usepackage{cite}

\newcommand{\fourier}[1]{\mathcal{F} \biggl[#1 \biggl]}
\newcommand{\ifourier}[1]{\mathcal{F}^{-1} \biggl[#1 \biggl]}
\newcommand{\sfourier}[1]{\mathcal{F} \bigl[#1 \bigr]}
\newcommand{\sifourier}[1]{\mathcal{F}^{-1} \bigl[#1 \bigr]}

\newcommand{\fft}[1]{\mathcal{{\rm FFT}} \biggl[#1\biggl]}
\newcommand{\ifft}[1]{\mathcal{{\rm FFT^{-1}}} \biggl[#1\biggl]}
\newcommand{\sfft}[1]{\mathcal{{\rm FFT}} [#1]}
\newcommand{\sifft}[1]{\mathcal{{\rm FFT^{-1}}} [#1]}

\newcommand{\nufftone}[1]{\mathcal{{\rm NUFFT_1}} \biggl[#1\biggl]}
\newcommand{\nuffttwo}[1]{\mathcal{{\rm NUFFT_2}} \biggl[#1\biggl]}
\newcommand{\snufftone}[1]{\mathcal{{\rm NUFFT_1}} [#1]}

\newcommand{\nuasmone}[1]{\mathcal{{\rm NUASM_1}} \biggl[#1\biggl]}
\newcommand{\nuasmtwo}[1]{\mathcal{{\rm NUASM_2}} \biggl[#1\biggl]}

\newcommand{\eq}[1]{Eq.(\ref{#1})}

\begin{document}

\twocolumn[ 

\title{Non-uniform sampled scalar diffraction calculation \\
using non-uniform fast Fourier transform}

\author{Tomoyoshi Shimobaba$^1$, Takashi Kakue$^1$, Minoru Oikawa$^1$, Naohisa Okada$^1$,  \\
Yutaka Endo$^1$, Ryuji Hirayama$^1$, and Tomoyoshi Ito$^1$}

\address{
$^1$Graduate School of Engineering, Chiba University, 1-33 Yayoi-cho, Inage-ku, Chiba 263-8522, Japan\\
$^*$Corresponding author: shimobaba@faculty.chiba-u.jp
}

\begin{abstract}
Scalar diffraction calculations such as the angular spectrum method (ASM) and Fresnel diffraction, are widely used in the research fields of optics, X-rays, electron beams, and ultrasonics.
It is possible to accelerate the calculation using fast Fourier transform (FFT); unfortunately, acceleration of the calculation of non-uniform sampled planes is limited due to the property of the FFT that imposes uniform sampling.
In addition, it gives rise to wasteful sampling data if we calculate a plane having locally low and high spatial frequencies. 
In this paper, we developed non-uniform sampled ASM and Fresnel diffraction to improve the problem using the non-uniform FFT.
\end{abstract}
\ocis{
(0070) Fourier optics and signal processing; 
(090.1760) Computer holography; 
(090.2870) Holographic display; 
(090.5694) Real-time holography; 
(090.1995) Digital holography.}
] 

\noindent Scalar diffraction calculations such as the angular spectrum method (ASM) and Fresnel diffraction \cite{goodman} are major methods used in wide-ranging optics, especially, important tools for computer-generated holograms (CGHs), digital holography, the design of diffractive optical elements, the analysis of beam propagation and so forth.
Generally, these diffractions are expressed as follows:
\begin{eqnarray}
u_2(\bm x_2) &=&  \int u_1(\bm x_1)  h_z(\bm x_2 - \bm x_1) d {\bm x_1} \nonumber  \\ 
&=& \ifourier{ \fourier{u_1(\bm x_1)} H_z(\bm f) }, 
\label{eqn:scalar}
\end{eqnarray}
where $\sfourier{\cdot}$ and $\sifourier{\cdot}$ are forward and inverse Fourier transforms, $u_1(\bm x_1)$ and $u_2(\bm x_2)$ are the source and destination planes, $\bm f=(f_x,f_y)$ is the position vector in the Fourier domain, and $z$ is the propagation distance between the source and destination planes, $h_z(\bm x_2 - \bm x_1)$ is the impulse response, and $H_z(\bm f)=\sfourier{h_z(\bm x_1)}$ is the transfer function. 

ASM is expressed as the following equations:
\begin{eqnarray}
u_2(\bm x_2) &=&  \int \fourier{u_1(\bm x_1)}  \exp \biggl( 2 \pi i z \sqrt{\frac{1}{\lambda^2} - |\bm f|^2} ~ \biggr)  \times \nonumber  \\ 
&& ~~~~~~~~~~~~~ \exp(2\pi i {\bm f \bm x_2}) d {\bm f} \nonumber  \\ 
&=& \ifourier{ \fourier{u_1(\bm x_1)} H_z(\bm f) }, 
\label{eqn:asm}
\end{eqnarray}
where $H_z(\bm f)= \exp ( 2 \pi i z \sqrt{\frac{1}{\lambda^2} - |\bm f|^2} )$ is the transfer function of ASM. 
In numerical calculation, it is possible to accelerate the calculation using  fast Fourier transform (FFT) as follows:
\begin{equation}
u_2(\bm m_2) = \ifft{ \fft{u_1(\bm m_1)} H_{z,\Delta_f}(\bm m_f) }, 
\label{eqn:asm_fft}
\end{equation}
where,  $\bm m_1=(m_1, n_1)$ and $\bm m_2=(m_2, n_2)$ are the position vector on the spatial domain as integer.
$\bm m_f =(m_f, n_f)$ is the position vector on the frequency domain as integer.
The source, destination planes and  frequency domain are sampled by the rates of $\Delta_1, \Delta_2$ and $\Delta_f$, respectively.
The operators $\sfft{\cdot}$ and $\sifft{\cdot}$ are forward and inverse FFTs, respectively.
The transfer function $H_{z,\Delta}(\bm m_f)$ \cite{shift_asm} is defined by, 
\begin{eqnarray}
H_{z, \Delta}(\bm m_f)&=& \exp \left( 2 \pi i \left(\bm o \bm m_f + z \sqrt{\frac{1}{\lambda^2} - |\bm m_f \Delta|^2} \right) \right) \times \nonumber \\
&& {\rm Rect}(\frac{m_f - c_m}{w_m}, \frac{n_f - c_n}{w_n})
\label{eqn:transfer}
\end{eqnarray}
where $\Delta$ is the sampling rate, and the offset parameter $\bm o=(o_x,o_y)$ is for off-axis calculation.
The two-dimensional rectangle function $\rm Rect(\cdot)$ is capable of calculating long propagation of ASM, respectively. 
$(w_m,w_n)$ and $(c_m,c_n)$ are the band-widths and the center of the band widths, respectively.
See Ref.\cite{shift_asm} for the determination of these parameters.
Likewise, we can calculate Fresnel diffraction using the following transfer function:
\begin{equation}
H_{z, \Delta}(\bm m_f)=\fft{\frac{\exp(ikz) \exp(i \frac{(\bm m_1 \Delta+\bm o)^2}{\lambda z})}{i \lambda z} }. 
\label{eqn:transfer_fre}
\end{equation}

Note that we have to expand the calculation size by zero-padding to $N' \times N'$ where $N'=2N$ ($N$ is the horizontal and vertical pixel numbers of planes) in order to avoid wraparound by the circular convolution of \eq{eqn:asm_fft}.
Therefore, the sampling rate $\Delta$ in the transfer functions is $\Delta=1/(N' \Delta_1)$.

However, using FFT imposes the same sampling rates on the source and destination planes ($\Delta_1 = \Delta_2$).
In order to solve this restriction, scaled-Fresnel diffraction \cite{sfre_a, sfre1,sfre2, sfre_b,sfre3} and scaled-ASMs \cite{sasm1,sasm2,sasm3} that can address different sampling rates on source and destination planes were proposed.
In Refs. \cite{sfre_a, sfre1,sfre2,sfre_b,sfre3, sasm1,sasm3}, chirp-z transform (a.k.a. scaled Fourier transform) is used instead of normal FFT.
In Ref. \cite{sasm2}, non-uniform sampling FFT (NUFFT) \cite{nufft4} is used.
The scaling operation is, for example, useful for lensless zoomable holographic projection \cite{zoom,zoom2}, digital holographic microscopy \cite{dhm} and fast CGH calculation \cite{cgh}.
Even though scaled-ASMs can address different sampling rates on source and destination planes, each plane still requires uniform-sampling.

In this paper, we develop non-uniform sampled ASM (NU-ASM) and Fresnel diffraction (NU-FRE) to overcome the restriction using the non-uniform FFT.
The uniform-sampled scalar diffractions give rise to wasteful sampling data if we calculate a plane having locally low and high spatial frequencies. 
In contrast, NU-ASM and NU-FRE are expected to be useful for the  situation.
NU-ASM is a generalized version of the scaled-ASM of Ref. \cite{sasm2}.

First, we consider NU-ASM.
We need three types of NU-ASM, that is, the first NU-ASM$_1$ calculates ASM from a non-uniform sampled source plane to the uniform-sampled destination plane.
The second NU-ASM$_2$ is the opposite version of the first.
The third, NU-ASM$_3$, calculates ASM on both non-uniform sampled source and destination planes.

By applying NUFFT to \eq{eqn:asm_fft} to a source plane, we can straightforwardly obtain NU-ASM$_1$:
\begin{eqnarray}
u_2(\bm m_2) &=& \nuasmone{ u_1(\bm x_1)} \nonumber \\
&=& \ifft{ \nufftone{u_1(\bm x_1)}  H_{z,\Delta}(\bm m_f) },  \nonumber \\
\label{eqn:nuasm1}
\end{eqnarray}
where $\snufftone{\cdot}$ is the type 1 of NUFFT, $\Delta=1/(N'\Delta_2)$ and $u_1(\bm x_1)$ is the non-uniform sampled source plane.
Several implementations of NUFFT were proposed.
We used Greengard and Lee's NUFFT \cite{nufft4}.
Two types of NUFFT are defined.
NUFFT type 1 is defined as follows:
\begin{equation}
F(\bm m_f) = \nufftone{f(\bm x_1')} = \sum_{\bm x_1'} f(\bm x_1') \exp(- i \pi \bm m_f \bm x_1'),
\label{eqn:nufft1}
\end{equation}
where, $\bm m_f$ is uniform sampled with the integer value on the frequency domain.
Practically, the right-side of the equation is accelerated by FFT, gridding algorithm and deconvolution \cite{nufft4}. 

\begin{figure}[htb]
\centerline{
\includegraphics[width=9.5cm]{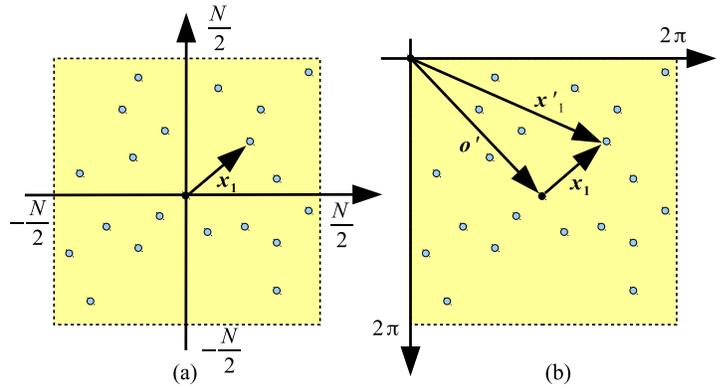}}
\caption{Coordinate conversion for NUFFT. (a) coordinate system of the original plane (b) coordinate system of the plane for NUFFT. Blue circles indicate non-uniform sampled points.}
\label{fig:fig_convert}
\end{figure}

Due to the NUFFT property, we need to normalize the coordinate of the source plane as shown in Fig.\ref{fig:fig_convert}.
The original coordinate system of the source plane $\bm x_1$ is defined as $\bm x_1 \in [-N/2,N/2) \times [-N/2,N/2)$ as shown in Fig.\ref{fig:fig_convert}(a). 
The NUFFT requires the range of the coordinate, $\bm x'_1 \in [0, 2\pi) \times [0, 2\pi)$  as shown in Fig.\ref{fig:fig_convert}(b).
Therefore, we convert the coordinate system by $\bm x'_1 = 2 \pi (\bm o' + \bm x_1) / N$ where $\bm o'=(N/2,N/2)$.

NUFFT type 2 is defined as follows:
\begin{equation}
f(\bm x_2') = \nuffttwo{F(\bm m_f)} = \sum_{\bm m_f} F(\bm m_f) \exp(i \pi \bm m_f \bm x_2'),
\label{eqn:nufft2}
\end{equation}
where $f(\bm x_2')$ is the non-uniform sampled plane.
Note that we need to normalize the coordinate of the destination plane for the same reason as that of  NUFFT$_1$.
For more details, see \cite{nufft4}.

Likewise, NU-ASM$_2$ is obtained as follows:
\begin{eqnarray}
u_2(\bm x_2)&=&\nuasmtwo{u_1(\bm m_1)}, \nonumber \\
&=&\nuffttwo{ \fft{u_1(\bm m_1)}  H_{z,\Delta}(\bm m_f) }, \nonumber \\
\label{eqn:nuasm2}
\end{eqnarray}
where $\Delta=1/(N'\Delta_1)$.
$\bm m_1$ and $\bm m_f$ are uniform sampled.
$\bm x_2$ is non-uniform sampled and the range is $\bm x_2 \in [-N/2,N/2) \times [-N/2,N/2)$.
The NUFFT$_2$ requires the range of the coordinate, $\bm x'_2 \in [0, 2\pi) \times [0, 2\pi)$.
Therefore, we convert the coordinate system by $\bm x'_2 = 2 \pi (\bm o' + \bm x_2) / N$.

We obtain NU-ASM$_3$ by combining NU-ASM$_1$ and NU-ASM$_2$.
In addition, using \eq{eqn:transfer_fre} instead of the transfer functions of NU-ASM, we can straightforwardly derive three types of NU-FRE, that is, NU-FRE$_1$, NU-FRE$_2$ and NU-FRE$_3$ .

Let us compare the NU-ASM and NU-FRE with Rayleigh-Sommerfeld (RS) diffraction.
RS diffraction is rigorous scalar diffraction and is expressed as,
\begin{equation}
u_2({\bm x_2})= \frac{1}{2 \pi} \int \!\! \int  u_1(\bm x_1) \left\{ \frac{z}{r} (1-ikr) \frac{\exp(i k r)}{r^2} \right\}  d{\bm x_1},
\label{eqn:huy_diff}
\end{equation}
where $k$ is the wave number and $r=\sqrt{|\bm x_1 - \bm x_2|^2 + z^2}$.
RS diffraction with non-uniform sampled planes is calculated by directly numerical integration, which has the complexity of $O(N^4)$.

\begin{figure}[htb]
\centerline{
\includegraphics[width=8cm]{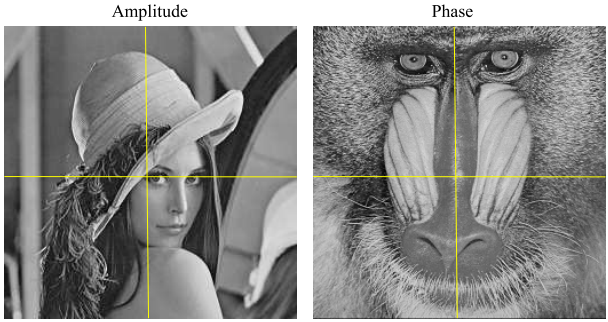}}
\caption{Source plane with $256 \times 256$ pixels. ``Lenna'' as the amplitude distribution of the source plane, and the image ``Mandrill'' as the phase distribution where the pixel value 0 corresponds to the phase value $-\pi/2$ and the pixel value 255 corresponds to the phase value $+\pi/2$.}
\label{fig:fig_lena}
\end{figure}

We use Fig.\ref{fig:fig_lena} as the non-uniform sampled source plane with $256 \times 256$ pixels.
The image ``Lenna'' is the amplitude distribution of the source plane, and the image ``Mandrill'' is the phase distribution where the pixel value 0 corresponds to the phase value $-\pi/2$ and the pixel value 255 corresponds to the phase value $+\pi/2$.
The calculation condition is as follows: the wavelength is 500 nm and the sampling rate on the uniform-sampled destination plane is $\Delta_2=5\mu$m.
The source plane is non-uniform-sampled, where the sampling rates on the first to fourth quadrants are $0.9\Delta_2, 0.8\Delta_2, 0.7\Delta_2$ and $\Delta_2$, respectively.

\begin{figure}[htb]
\centerline{
\includegraphics[width=9cm]{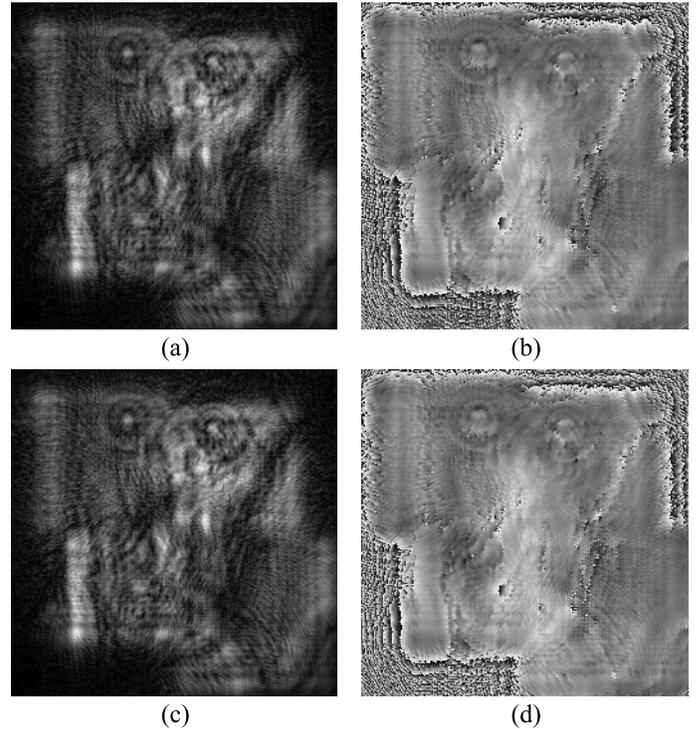}}
\caption{Diffracted results of the amplitude and phase distribution by RS diffraction and NU-ASM. (a) amplitude by RS diffraction (b) phase by RS diffraction (c) amplitude by NU-ASM (d) phase by NU-ASM.}
\label{fig:fig_rs_nu}
\end{figure}

\begin{figure}[htb]
\centerline{
\includegraphics[width=9cm]{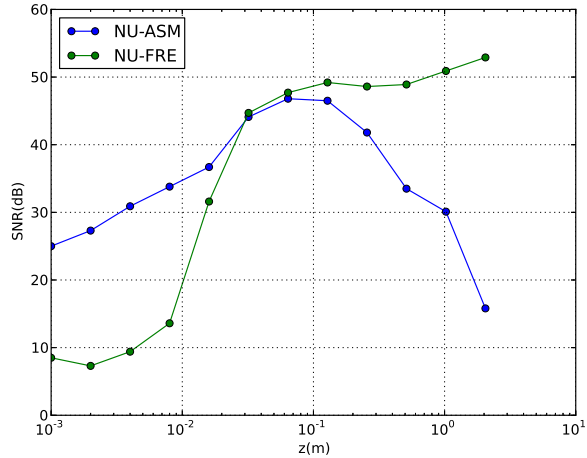}}
\caption{Errors of NU-ASM$_1$ and NU-FRE$_1$ to RS diffraction by  the signal-to-noise ratio (SNR).}
\label{fig:fig_psnr}
\end{figure}

Figure \ref{fig:fig_rs_nu}(a) and (b) are the amplitude and phase on the diffracted result by RS diffraction at $z=8$mm.
Figure \ref{fig:fig_rs_nu}(c) and (d) are the amplitude and phase on the diffracted result by NU-ASM$_1$.
In Fig. \ref{fig:fig_psnr},  we measure the error of NU-ASM$_1$ and NU-FRE$_1$ to RS diffraction, which is the criteria, by the signal-to-noise ratio (SNR).
At short distance propagation, the SNR of NU-ASM$_1$ is good quality, while that of NU-FRE$_1$ is poor quality.
Whereas, at long distance propagation, the SNR of NU-ASM$_1$ is decreased because the band-limited transfer function of \eq{eqn:transfer} has a small area, which is pointed out by Ref.\cite{basm}.
While, the SNR of NU-FRE$_1$ is increased.
The average calculation time of the NU-ASM$_1$ is about 46 ms, while that of RS diffraction is about 76 s. 
We use an Intel Core i7-2600S CPU and eight CPU threads, and multi-thread version of FFTW \cite{fftw} as the FFT library.

\begin{figure}[htb]
\centerline{
\includegraphics[width=9cm]{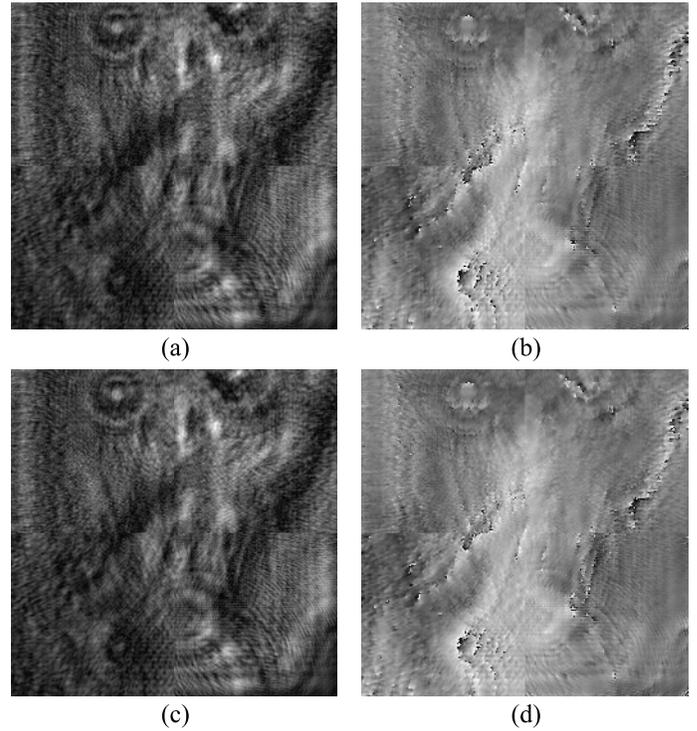}}
\caption{Diffracted results of the amplitude and phase distribution by RS diffraction and NU-ASM. (a) amplitude by RS diffraction (b) phase by RS diffraction (c) amplitude by NU-ASM (d) phase by NU-ASM.}
\label{fig:fig_rs_nu2}
\end{figure}

Figure \ref{fig:fig_rs_nu2} shows diffracted results from a uniform-sampled source plane to non-uniform-sampled destination plane by RS diffraction and NU-ASM$_2$.
We use Fig.\ref{fig:fig_lena} as the uniform-sampled source plane at the sampling rate $\Delta_1=5 \mu$m.
The destination plane is non-uniform-sampled, where the sampling rates on the first to fourth quadrants are $0.9\Delta_1, 0.8\Delta_1, 0.7\Delta_1$ and $\Delta_1$, respectively.
Figure \ref{fig:fig_rs_nu2}(a) and (b) are the amplitude and phase on the diffracted result by RS diffraction at $z=8$mm.
Figure \ref{fig:fig_rs_nu2}(c) and (d) are the amplitude and phase on the diffracted result by NU-ASM$_2$.
The SNR is maintained at approximately 32 dB.

We conclude this work.
This paper proposes the non-uniform sampled scalar diffractions, NU-ASM and NU-FRE, based on NUFFT.
These diffractions are categorized into three types: type 1 is the diffraction from a non-uniform sampled source plane to a uniform sampled destination plane, the type 2 is the opposite situation of type 1; and type 3 is the diffraction between a non-uniform sampled source and destination planes.
NU-ASM is suitable for the short propagation, while NU-FRE are suitable for the long propagation.
In addition, the calculation times of NU-ASM and NU-FRE is faster than RS diffraction.
In future, we will attempt to apply the non-uniform sampled scalar diffractions to CGH calculation, the design of optical elements, and optical encryption and watermarking. 
These diffraction calculations will be provided in our open-source library: CWO++ library \cite{cwo}.   


This work is supported by the Ministry of Internal Affairs and Communications, Strategic Information and Communications R\&D Promotion Programme (SCOPE)(09150542), Japan Society for the Promotion of Science (JSPS) KAKENHI (Young Scientists (B) 23700103) 2011, and the NAKAJIMA FOUNDATION.


\end{document}